Arrested kinetics of the CO-I to FM-M transition and the phase separated glassy state in manganites: a Comment on **Kundhikanjana et al PRL 115 (2015) 265701** and **Kundhikanjana et al arXiv:1306.3065v1 (2013)**

P Chaddah*
UGC-DAE Consortium for Scientific Research,
Indore 452001
India.

The coexistence of competing phases (referred to as PS state in the papers [1,2] being commented on) first gained prominence in the half-doped manganites, where these were observed to persist on cooling below the first-order phase transition temperature. One of these phases is in a metastable state, and is persisting to the lowest temperature because the glass-like arrest of kinetics has interrupted the phase transition. The other phase, created before the transition was interrupted, is in equilibrium. The metastable phase would show relaxation behavior that becomes slower as temperature falls, in contrast to supercooled states where relaxation rate rises as temperature falls [3]. Since the first order transition between the charge-ordered insulator (CO-I) and the ferromagnetic metal (FM-M) is also driven by magnetic field, similar phase separated coexistence is seen on varying H but the dependence of relaxation rates on magnetic field is not unique in that it depends on what the low temperature equilibrium phase is. The acceptance of this 'interrupted phase transition' explanation of phase coexistence resulted after observations of predictions under novel variations in (H, T). Most of the observations were of macroscopic properties, while a few observations were at mesoscopic and microscopic scales. The papers being commented on do report a few such macroscopic measurements, but are silent about the earlier detailed predictions and measurements under novel thermo-magnetic histories. These two papers add new mesoscopic measurement techniques, but need to perform measurements under novel thermo-magnetic histories if they have to support (or negate) the 'interrupted phase transition' explanation of phase coexistence. We discuss this briefly below.

The interrupted phase transition explanation (called "kinetic arrest") was first put forward based on data in a ferro- to antiferro-magnetic transition in doped $CeRu_2$ with similarities to the half-doped manganites clearly highlighted [4]. As this explanation was pursued in other materials showing similar first-order magnetic transitions, concepts and tests were developed under novel paths in (H, T) space and it was only then applied to the hotly studied manganites [5]. Further predictions of behavior under novel thermo-magnetic histories were developed, and the protocol of CHUF (Cooling and Heating in Unequal Fields) was found to give visually striking results [6,7]. A brief, but up-to-date, review of these and subsequent developments is available in reference [8].

There are three indicators of kinetic arrest (or the 'interrupted phase transition' explanation) that manifest in various measurements [8].

The first is in conventional measurements where the first order magnetic transition, between two states of competing free energies, is observed as a sharp large change in a macroscopic physical property with lowering temperature. The temperature at which the transition occurs ($T_C$) can be reduced by varying the magnetic field in which the sample is cooled. As $T_C$ is lowered the magnitude of this sharp change is reduced, and the change may eventually be totally inhibited. This anomaly is reported in many materials and is observed with decreasing H when the transition is to a ferromagnetic state with lowering temperature, and with increasing H when the transition is to an anti-ferromagnetic state with lowering temperature [8]. Such studies are not reported in the two papers being commented on.

The second corresponds to conventional measurements of the macroscopic physical property under isothermal variation of magnetic field. This would also show a sharp large change both while increasing and while decreasing field. The transitions are normally seen at different values of field due to the hysteresis associated with supercooling and superheating. But when the PS state persists due to kinetic arrest, then devitrification overrules supercooling at low temperatures. And the field dependence of the devitrification temperature has a sign opposite to that of the field dependence of supercooling spinodal temperature. This produces a non-monotonic behavior with temperature, of one of the fields at which the transition occurs in an isothermal variation, and is the second qualitative anomaly [8]. This is observed in figure 1(d) of Kundhikanjana et al [1], but the discussion is innocent of it being a predicted feature of kinetic arrest (or the 'interrupted phase transition' explanation).

The third anomaly is observed in measurements under the novel thermo-magnetic histories called CHUF. A subset of these is performed when reporting ZFC-warming data, with different warming fields. These are visually drastic when the low-temperature phase is ferromagnetic and the high-temperature phase is anti-ferromagnetic. The ZFC state is a kinetically arrested glasslike state, and warming results in the analogue of 'devitrification' of this glasslike state, followed at higher temperature by the equivalent of the melting transition. The dependence of the two transition temperatures, on the warming field, should have opposite signs for the case of kinetic arrest (or the 'interrupted phase transition' explanation). This is reported in figure 1(b) of reference [2], which is again innocent of exactly similar earlier data in figure 1 of reference [9] amply supported by mesoscopic MFM data. More detailed CHUF measurements of macroscopic properties have been reported in many materials as reviewed in reference [8], and these have been put on a firm basis with microscopic neutron diffraction measurements [10].

We strongly suggest that microwave impedance microscopy (MIM) studies on PCSMO, first reported by Kundhikanjana et al [2] in 2013, should now be performed under the CHUF protocol. It is to be noted that studies under the CHUF protocol have also helped establish that kinetic arrest of the first-order transition leads to the observed magnetic-glass state in $Mn_2PtGa$ [11]. Research supporting or negating earlier concepts is incremental to existing knowledge.

*Since retired. Email: chaddah.praveen@gmail.com

## References

1. W Kundhikanjana et al **Phys Rev Lett 115** (2015) 265701.
2. W Kundhikanjana et al **arXiv:1306.3065v1** (2013).
3. P Chaddah, K Kumar and A Banerjee **Phys Rev B 77** (2008) 100402.
4. M Manekar et al, **Phys Rev B 64** (2001) 104416.
5. K Kumar et al **Phys Rev B 73** (2006) 184435; A. Banerjee et al **J. Phys.: Condens. Matter 18** (2006) L605.
6. A. Banerjee, K. Kumar and P. Chaddah, **J. Phys.: Condens. Matter 21** (2009) 026002.
7. T. Sarkar, V. Pralong, and B. Raveau, **Phys.Rev.B 83** (2011) 214428; **Phys.Rev.B 84** (2011) 059904(E).
8. P Chaddah, **AIP Conference Proceedings 1661** (2015) 030002; **arXiv:1410.3254** (2014).
9. A Lakhani et al **J. Phys.: Condens. Matter 22** (2010) 032101.
10. V Siruguri et al **J. Phys.: Condens. Matter 25** (2013) 496011.
11. A Nayak et al, **arXiv:1304.4459** (2013); A Nayak et al **Phys Rev Lett 110** (2013) 127204.